\documentclass[onecolumn,sort&compress,numbers]{els-mrw} 

\usepackage{amsmath,amssymb,amsfonts,amsthm,makeidx,graphicx}
\usepackage{txfonts}
\usepackage{helvet}
\usepackage[colorlinks,citecolor=black,linktoc=all,linkcolor=black]{hyperref}
\usepackage{wasysym}

\begin{document}
	
	
	\chapter{Solar neutrinos}\label{chap1}
	
	\author[1]{Shaomin Chen}%
	\author[2]{Xun-Jie Xu}%

	\address[1]{\orgdiv{Department of Engineering Physics}, \orgname{Tsinghua University},  \orgaddress{Beijing 100084, China}}
	\address[2]{\orgname{Institute of High Energy Physics, Chinese Academy of Sciences}, \orgaddress{Beijing 100049, China}}
	

	\maketitle
	
	\begin{abstract}[Abstract]
Solar neutrinos, generated abundantly by thermonuclear reactions in the solar interior, offer a unique tool for studying astrophysics and particle physics. The observation of solar neutrinos has led to the discovery of neutrino oscillation, a topic currently under active research, and it has been recognized by two Nobel Prizes. In this pedagogical introduction to solar neutrino physics, we will guide
readers through several key questions: How are solar neutrinos produced? How are they detected? What is the solar neutrino problem, and how is it resolved by neutrino oscillation? This article also presents a brief overview of the theory of solar neutrino oscillation, the experimental achievements, new physics relevant to solar neutrinos, and the prospects in this field. 
\end{abstract}
	
	\begin{keywords}
		Solar neutrinos \sep Neutrino oscillation \sep MSW effect \sep Neutrino detection \sep Solar neutrino problem 
	\end{keywords}

	\section*{Objectives}
	\begin{itemize}
		\item Section \ref{sec:SSM} introduces the standard solar model, explains how neutrinos are produced in the Sun via thermonuclear reactions (pp chain and CNO cycles), and summarizes the predicted fluxes and energy spectra of solar neutrinos.
        \item Section \ref{sec:exp} introduces the detection of solar neutrinos, the history of solar neutrino experiments, and the well-known solar neutrino problem.
		\item Section \ref{sec:osc} introduces the standard theory of solar neutrino oscillation including the MSW effect and the day-night difference. 
		\item Section \ref{sec:exp-progress} reports on experimental achievements and progress in measuring the expected solar neutrino components from all the fusion processes.
        \item Section \ref{sec:beyond} discusses a few new physics effects relevant to solar neutrinos and how they can be probed by solar neutrino experiments. 
        \item Section \ref{sec:summary} summarizes the key facts about solar neutrinos and presents the outlook.
	\end{itemize}


\section{Introduction \label{sec:intro}}

Solar neutrinos represent one of the most enduring and exciting fields of research in astrophysics and particle physics.
From the astrophysical perspective, they offer a unique tool that enables us to gain direct insights into the solar interior. 
From the perspective of particle physics, the Sun serves as an intense source of neutrinos, allowing us to explore their fundamental properties and deepen our understanding of the underlying theories that govern these elusive particles.

Pioneered by Raymond Davis Jr. in the 1960s, the first solar neutrino experiment observed a deficit of solar neutrinos compared to theoretical predictions, providing the first hint of neutrino oscillation, which, among various other explanations for the deficit at that time, was ultimately confirmed as the only resolution of the solar neutrino problem by subsequent  experiments. 

The confirmation of neutrino oscillation has a profound impact on particle physics, as it implies that neutrinos---contrary to predictions made by the Standard Model (SM)---have nonzero masses. This discovery opens new avenues for exploring new physics beyond the SM. 
Thus, the study of solar neutrinos stands as one of the most successful examples demonstrating how new observations into the sky can lead to groundbreaking discoveries in fundamental physics.

In this article, we aim at presenting an pedagogical introduction to solar neutrino physics. In Secs.~\ref{sec:SSM} and \ref{sec:exp},  we will guide readers through a few questions including how neutrinos are produced from the Sun, how they are detected, why there was the deficit and how it was resolved by neutrino oscillation. We further present a brief yet self-contained formalism to facilitate quick understanding of the calculation of solar neutrino oscillation (Sec.~\ref{sec:osc}), and an overview of experimental achievements and progress (Sec.~\ref{sec:exp-progress}). We also briefly comment on open issues and new physics beyond the established framework. For a more comprehensive review, we refer to Ref.~\cite{Xu:2022wcq}.

\section{Standard solar model and solar neutrinos \label{sec:SSM}}

The Sun is a powerful source of neutrinos which are abundantly produced by the thermonuclear reactions inside.
The production of solar neutrinos is predicted by the Standard Solar Model (SSM), a theoretical framework crucial for understanding the structure and behavior of the Sun.
Table~\ref{tab:sun_parameters} lists a few well-known quantities of the solar profile. Among them, some are determined by observations (e.g.~luminosity, radius, mass, surface temperature) while others are predictions of the SSM, including solar neutrino fluxes.

\begin{table}[h]
	\centering
	\TBL{\caption{\label{tab:sun_parameters} A basic profile of the Sun. Some quantities are determined by observations, while others are predicted by the SSM.}
	}{
	\begin{tabular}{llll}
		\toprule 
  Parameter (observed) & Value &  Parameter (predicted) & Value\tabularnewline
  \midrule
  Luminosity & $3.828\times10^{26}$ W &  Central temperature & $1.54\times10^7$ K \\
  Radius     & $6.961\times10^{7}$ km &  Central density & $149~\text{g}~\text{cm}^{-3}$\\
  Mass       & $1.988\times10^{30}$ kg  &  Central pressure & $2.3\times 10^{16}~\text{Pa}$\\
  Surface temperature & $5.78\times10^3$ K &  Neutrino fluxes &  see Tab.~\ref{tab:flux}\\
  \bottomrule
	\end{tabular}
	}{}
\end{table}

The SSM is constructed upon principles of hydrostatic equilibrium, energy transport mechanisms, and the nuclear reactions that power the Sun's energy output. By integrating these principles with observational constraints such as solar radius, luminosity, age, elemental composition, and radiative opacity, detailed predictions can be made about the internal solar structure, including density, temperature, pressure, and neutrino fluxes. 

Including modern knowledge of neutrinos, the SSM is rather successful in predicting and correlating various observables of the Sun. Historically, the SSM had a long-standing problem: the observed solar neutrino flux was significantly lower than the prediction. 
This problem, known as the {\it solar neutrino missing problem},  was eventually resolved by neutrino oscillations. In what follows, we will explain how solar neutrinos are produced in the Sun. 

\subsection{Thermonuclear reactions in the Sun}

The solar core is mainly made up of hydrogen (about 74\% by mass), helium (about 24\%), and small amounts (less than 2\%) of heavier elements like oxygen, carbon, neon, and iron. At the core of the Sun, the density reaches around 150 grams per cubic centimeter, with temperatures soaring to about 15 million Kelvin. These extreme conditions allow the penetration of the Coulomb barrier between ions through the quantum tunneling effect, enabling thermonuclear reactions that convert hydrogen into helium through the proton-proton (pp) chain and the carbon-nitrogen-oxygen (CNO) cycle. 
In the Sun, the pp chain is responsible for 99\% of the total solar energy production, while the remaining  $\sim 1\%$ is produced by the CNO cycle\footnote{Despite its sub-dominance in solar energy production, the CNO cycle plays a more significant role in massive stars. For stars with masses above 1.3 times the solar mass, the CNO cycle dominates the energy production. }.

\begin{figure}
	\centering 
	\includegraphics[width=0.99\columnwidth]{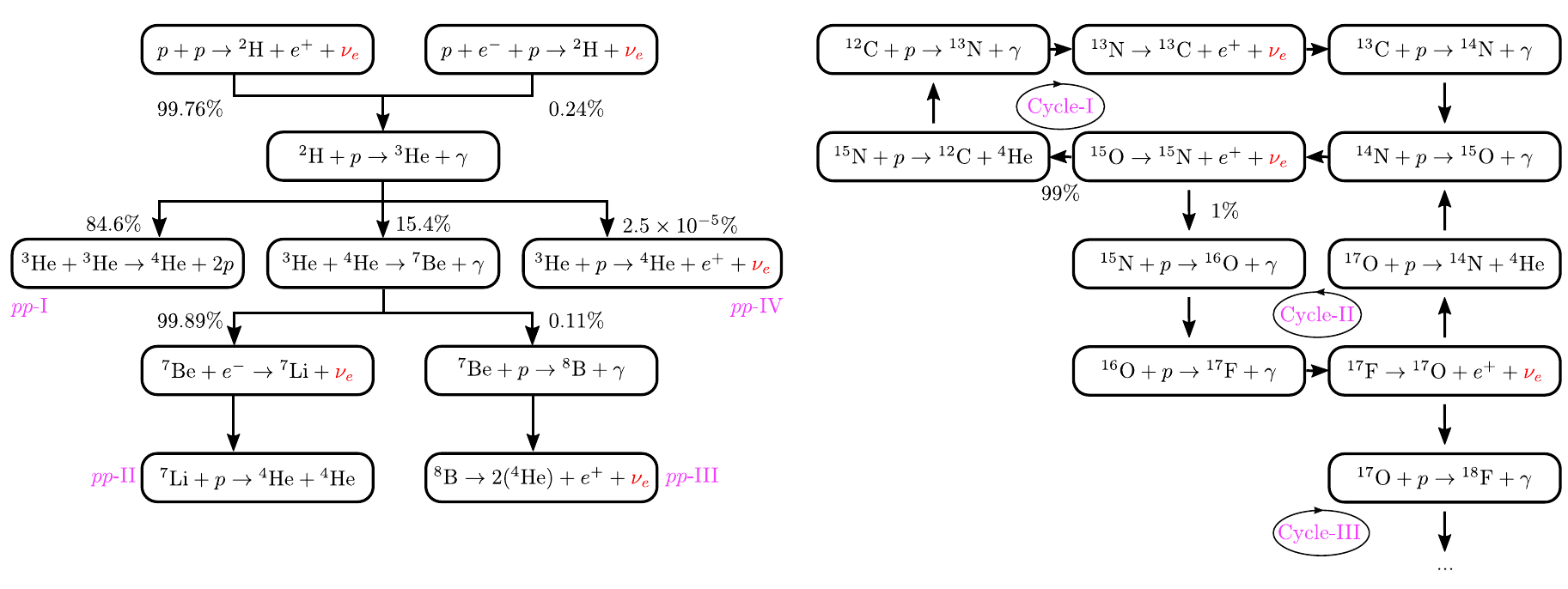}
	\caption{The pp chain (left) and the CNO cycle (right). In the pp chain, solar neutrinos are produced via five nuclear reactions and referred to as pp, pep, hep, $^7$Be, $^8$B neutrinos. 	
		In the CNO cycle, solar neutrinos are produced from the decays of $^{13}$N, $^{15}$O, and $^{17}$F. 
		\label{fig:pp-CNO}}
\end{figure}

As is shown in Fig.~\ref{fig:pp-CNO}, the pp chain starts with proton-proton fusion ($p+p\to {}^2{\rm H}+e^+ +\nu_e$) or, at a much lower rate, proton-electron-proton fusion ($p+e^++p\to {}^2{\rm H} +\nu_e$), which is possible due to the Coulomb potential of protons capturing an electron. Both processes produce electron neutrinos ($\nu_e$), but $\nu_e$ produced from the latter is monochromatic and more energetic. 
After the initial step of fusion,  subsequent nuclear reactions proceed, eventually ending up with four possible sub-chains, denoted by pp-I to pp-IV in Fig.~\ref{fig:pp-CNO}. Except for pp-I, each of these sub-chains contains a reaction that can produce neutrinos. 
Solar neutrinos produced from these reactions in the pp chain are usually referred to as pp, pep, hep, ${}^{7}{\rm Be}$, and  ${}^{8}{\rm B}$ neutrinos.

The CNO cycle, as the name suggests, involves carbon, nitrogen, and oxygen participating a cycle of nuclear reactions---see the right panel of Fig.~\ref{fig:pp-CNO}. Strictly speaking, it is not just one completely closed cycle.  Instead, it contains multiple cycles coupled together,  allowing some of the nuclear elements to leave the dominant cycle and join another cycle involving heavier elements. For instance, when  $^{15}{\rm N}$ is produced by Cycle-I in Fig.~\ref{fig:pp-CNO} via  $^{15}{\rm O}\to {}^{15}{\rm N}+e^+ +\nu_e $  and then captures a proton, $99\%$ will be converted to ${}^{12}{\rm C}+{}^{4}{\rm He}$ while around $1\%$ will be converted to ${}^{16}{\rm O}+\gamma$. 
So  $^{15}{\rm N}$ at this stage has a small probability of joining  Cycle-II in Fig.~\ref{fig:pp-CNO}. The same situation can also occur for $^{17}{\rm O}$, which may leave Cycle-II and join Cycle-III. Nevertheless, those cycles involving heavier nuclear elements are less important to the production of solar energy and neutrinos. If we only consider Cycle-I and Cycle-II, there are three reactions producing neutrinos, from $^{13}{\rm N}\to {}^{13}{\rm C}+e^+ +\nu_e$, $^{15}{\rm O}\to {}^{15}{\rm N}+e^+ +\nu_e $, and  $^{17}{\rm F}\to {}^{17}{\rm O}+e^+ +\nu_e $. The corresponding neutrinos are referred to as $^{13}{\rm N}$, $^{15}{\rm O}$, and  $^{17}{\rm F}$ neutrinos, respectively.

Note that in the pp chain, no stable elements heavier than $^4{\rm He}$ can be produced. Although some elements such as $^7{\rm Li}$,  $^7{\rm Be}$, and  $^8{\rm B}$ are produced, they decay rapidly, implying that the pp chain burns hydrogen only into helium.
In the CNO cycle, heavy elements like carbon and nitrogen remain almost unchanged after a complete cycle of reactions, implying that they participate in the reactions as catalysts. The net effect of the CNO cycle is also to burn hydrogen into helium. 
Since both series of reactions only produce $^4{\rm He}$, the abundance of elements heavier than $^4{\rm He}$---known as the metallicity (in astrophysics, ``metals'' refers to elements heavier than $^4{\rm He}$)---remains rather stable in the Sun. 
These ``metals'' not only play the role of catalysts in thermonuclear reactions, but also affect the opacity of the Sun. 
Therefore, the metallicity of the Sun is one of the key parameters of the SSM.

\subsection{Solar neutrino fluxes and spectra}

By evaluating the thermonuclear reaction rates in the Sun, one can calculate solar neutrino fluxes. Starting from John N. Bahcall's pioneering work~\cite{Bahcall:1968hc},  the calculation of solar neutrino fluxes has been continuously revised and improved, not only because the quality and quantity of input data for the SSM have increased, but also due to the increasing computational power that allows for more sophisticated simulations (e.g., switching from 1D to 3D, using non-local thermodynamic equilibrium, etc.) to be incorporated into the SSM.

\begin{table*}
	\centering
	\TBL{\caption{\label{tab:flux} Calculated solar neutrino fluxes at the Earth.}
	}{
	\begin{tabular}{ccccc}
		\toprule 
		Flux [${\rm cm}^{-2}{\rm s}^{-1}$] & BSB05-GS98 & BSB05-AGS05 & B16-GS98 & B16-AGSS09\tabularnewline
		\midrule 
		$\Phi_{{\rm pp}}/10^{10}$ & $5.99(1\pm0.009)$ & $6.06(1\pm0.007)$ & $5.98(1\pm0.006)$ & $6.03(1\pm0.005)$\\[1mm]
		$\Phi_{{\rm pep}}/10^{8}$ & $1.42(1\pm0.015)$ & $1.45(1\pm0.011)$ & $1.44(1\pm0.01)$ & $1.46(1\pm0.009)$\\[1mm]
		$\Phi_{{\rm hep}}/10^{3}$ & $7.93(1\pm0.155)$ & $8.25(1\pm0.155)$ & $7.98(1\pm0.30)$ & $8.25(1\pm0.30)$\\[1mm]
		$\Phi_{^{7}{\rm Be}}/10^{9}$ & $4.84(1\pm0.105)$ & $4.34(1\pm0.093)$ & $4.93(1\pm0.06)$ & $4.50(1\pm0.06)$\\[1mm]
		$\Phi_{^{8}{\rm B}}/10^{6}$ & $5.69(1_{-0.147}^{+0.173})$ & $4.51(1_{-0.113}^{+0.127})$ & $5.46(1\pm0.12)$ & $4.50(1\pm0.12)$\\[1mm]
		$\Phi_{^{13}{\rm N}}/10^{8}$ & $3.05(1_{-0.268}^{+0.366})$ & $2.00(1_{-0.127}^{+0.145})$ & $2.78(1\pm0.15)$ & $2.04(1\pm0.14)$\\[1mm]
		$\Phi_{^{15}{\rm O}}/10^{8}$ & $2.31(1_{-0.272}^{+0.374})$ & $1.44(1_{-0.142}^{+0.165})$ & $2.05(1\pm0.17)$ & $1.44(1\pm0.16)$\\[1mm]
		$\Phi_{^{17}{\rm F}}/10^{6}$ & $5.83(1_{-0.420}^{+0.724})$ & $3.25(1_{-0.142}^{+0.166})$ & $5.29(1\pm0.20)$ & $3.26(1\pm0.18)$\\
		\bottomrule
	\end{tabular}
	}{}
\end{table*}

Table~\ref{tab:flux} presents solar neutrino fluxes obtained by the Bahcall-Serenelli-Basu~(BSB) calculation~\cite{Bahcall:2005va} and the Barcelona-2016~(B16) calculation~\cite{Vinyoles:2016djt} based on different SSM data sets including data sets named GS98, AGS05, AGSS09. These data sets are named by the initials of the authors conducting the calculations and the years of the publications. 
In general, solar neutrino fluxes computed by different groups based on different SSM data sets differ from each other.  The differences for pp are small, only at the percent level or less, as can be seen from Tab~\ref{tab:flux}. But for $^8$B and CNO neutrinos, the differences are significantly larger.

The most important factor behind these differences is solar metallicity. Currently, there are two competing classes of solar models: high-metallicity and low-metallicity models. High-metallicity models (such as GS98) predict higher $^8$B and CNO neutrino fluxes than low-metallicity models (such as AGS05 and AGSS09), not only because the abundance of the catalysts mentioned above is higher but also due to higher radiative opacity caused by the heavy elements. The opacity inhibits heat transfer via radiation and thus increases the core temperature, which in turn raises the nuclear reaction rates and the corresponding neutrino fluxes. 

The two classes of models have their respective problems, so which one can more accurately predict the solar neutrino fluxes is still unresolved. Recent calculations of solar neutrino fluxes usually consider both of them. Generally speaking, high-metallicity models are in better agreement with helioseismological observations\footnote{Helioseismology studies the interior of the Sun based on vibrations of the solar surface, similar to seismology for the Earth.}, while low-metallicity models are favored by more advanced simulations but are in tension with helioseismological data. This unresolved issue is known as the solar metallicity problem. We do not intend to expand further in this pedagogical article and refer interested readers to recent reviews~\cite{Gann:2021ndb,Xu:2022wcq}.

\begin{figure}
	\centering 
	\includegraphics[width=0.49\textwidth]{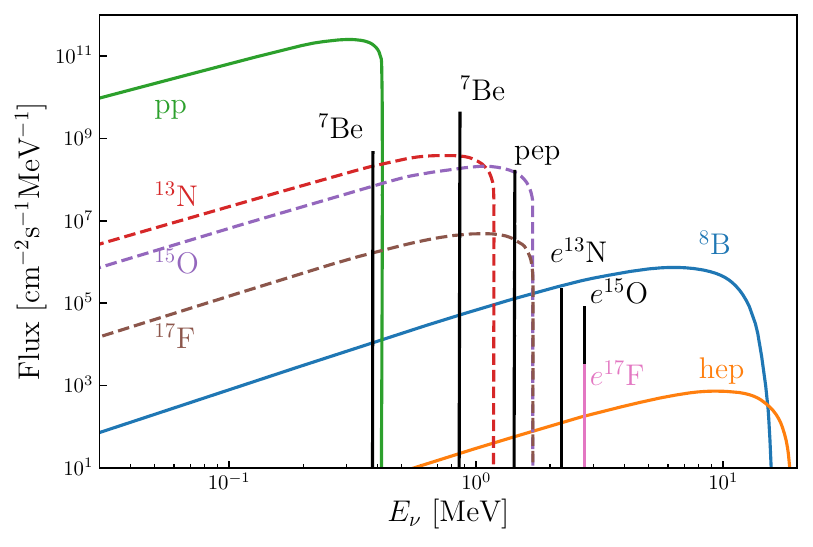} 
	\includegraphics[width=0.49\textwidth]{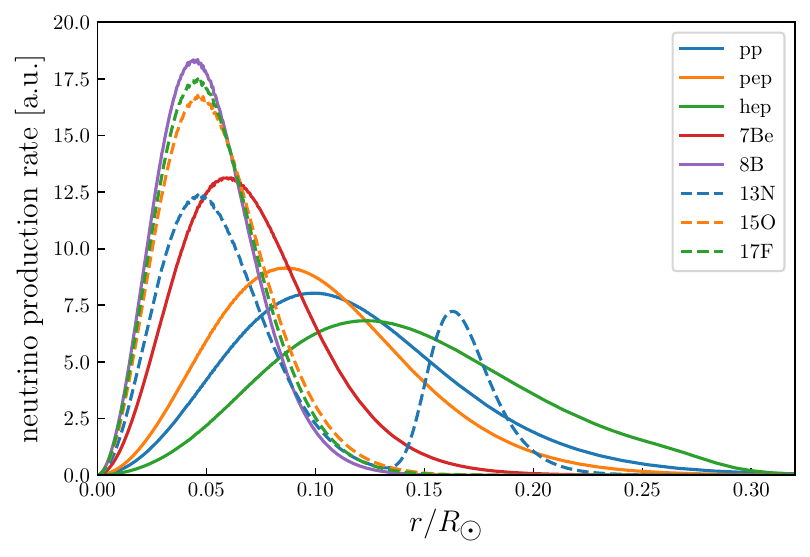} 
	\caption{Left: The energy spectra of solar neutrino fluxes. Note that monochromatic spectra are in units of ${\rm cm}^{-2}{\rm s}^{-1}$. Right:  The production rates of solar neutrinos  as a function of the radius $r$. \label{fig:flux_E_and_r}}
\end{figure}

The shapes of solar neutrino energy spectra are mainly determined by the kinematics of corresponding reaction processes and are almost independent of solar models. Unlike the total fluxes, which are affected by the uncertainties of solar models, the spectral shapes are invulnerable to potential variations in the core profile (e.g.~variations in temperature and density). This is because the energy released by a nuclear reaction is much higher than the kinetic energy of the initial-state particles in the reaction. The former is typically above MeV, while the latter is around the core temperature ($\sim$keV). Therefore, the spectral shapes in Bahcall's calculation~\cite{Bahcall1989} are still widely used in modern solar neutrino calculations. 
Figure~\ref{fig:flux_E_and_r} (left panel) shows the solar neutrino energy spectra obtained using Bahcall's spectral shapes and the total flux data of B16-GS98 in Tab.~\ref{tab:flux}. Also shown is the radial distribution of neutrino production in the Sun as a function of the radial distance $r$ divided by the solar radius $R_{\astrosun}$. 

As is shown in Fig.~\ref{fig:flux_E_and_r}, some of the solar neutrino energy spectra (such as pep, $^{7}$Be, etc.) are monochromatic. This is because the final states of these reactions only contain two particles. When the total energy of the initial states is fixed (which is approximately the case as all initial particles are non-relativistic), the two-body kinematics dedicates that the neutrinos are monochromatic. For $^{7}$Be neutrinos produced from ${}^{7}{\rm Be} +e^-\to {}^{7}{\rm Li} +\nu_e$, the $^{7}$Li nucleus may be in the ground state or an excited state, causing two monochromatic lines at 0.861 MeV (with a branching ratio of 90\%) and 0.383 MeV (10\%). In addition to the reactions in the pp chain, some reactions in the CNO cycle are accompanied by electron-capture processes, which also produce monochromatic neutrinos. For example,  ${}^{13}{\rm N}  \to {}^{13}{\rm C} + e^+ + \nu_e$  implies that ${}^{13}{\rm N} + e^-  \to {}^{13}{\rm C} + \nu_e$ is also possible, though at a much lower reaction rate. These fluxes are denoted by $e {}^{13}{\rm N}$, $e {}^{15}{\rm O}$, and $e {}^{17}{\rm F}$ in Fig.~\ref{fig:flux_E_and_r}.

\section{Solar neutrino detection: principles and methodologies \label{sec:exp}}
The first solar neutrino experiment was conducted by Davis at Homestake in the 1960s, using the radiochemical method to detect solar neutrinos. 
Since then, many experiments have been carried out based on diverse detection technologies, including radiochemical, Cherenkov, and liquid-scintillator detectors.

In the design of a solar neutrino experiment, two crucial factors must be considered. 

First, the selection of experimental sites is critical.  Solar neutrinos are at the same energy scale as various radioactive decays of unstable nuclear isotopes, which could be continuously produced by cosmic rays interacting with the detector if it is not well shielded. 
These radioactive decays of cosmogenic unstable isotopes could mimic solar neutrino events, creating a troublesome background for detection. To tackle this problem, solar neutrino experiments are usually conducted deep underground to minimize such interference.

Second, the choice of reaction processes and detection techniques is also important.  Neutrino capture processes may benefit from low reaction thresholds, but so far, have only been successfully applied to radiochemical detectors, which can not provide real-time measurements. Neutrino-electron scattering is currently the most important detection process for Cherenkov detectors. These detectors feature real-time measurements but need relatively high detection thresholds due to fewer Cherenkov photons emitted when the electron energy decreases.  Liquid-scintillator detectors can have lower thresholds but usually lose directionality.

\subsection{Radiochemical experiments}

Radiochemical experiments are based on neutrino-capture processes: $\nu_e + {{}^A_Z{\rm X}} \rightarrow ~{{}^A_{Z+1}{\rm Y}}+e^-$, 
where a solar neutrino $\nu_e$ is captured by a nucleus ${{}^A_Z{\rm X}}$ (here $A$ and $Z$ denote the numbers of nucleons and protons in the nucleus ${\rm X}$), 
resulting in the formation of a daughter nucleus 
${{}^A_{Z+1}{\rm Y}}$ and the emission of an electron. To identify the process in the experiment, the daughter nuclide needs to be radioactive, which means it is unstable and decays with a sufficiently long lifetime.  
After the target material being exposed to solar neutrinos for a certain period,  the produced daughter isotope is extracted via chemical methods. Then the number of these radioactive particles can be counted using proportional counters 
when they decay. Therefore, the lifetime of the daughter isotope needs to be reasonably long such that it does not decay instantly during the exposure step, and meanwhile can decay effectively during the second step after being chemically extracted. Note that such experiments cannot perform real-time measurements due to the two-step procedure mentioned above. They merely count rates of the neutrino-capture reactions over a designed period.

The first solar neutrino experiment, Davis's Homestake experiment, is based on this detection principle.
The experiment utilized 615 tons of liquid perchloroethylene \(\text{C}_2\text{Cl}_4\) which both $^{35}{\rm Cl}$ (76\%) and $^{37}\text{Cl}$  (24\%)  atoms but only the latter is responsible for neutrino capture: $\nu_e +{^{37}\text{Cl}} \rightarrow ~^{37}\text{Ar} + e^-$.
This reaction has an energy threshold of 814 keV. 
The daughter nucleus \(^{37}\text{Ar}\) has a half-life of 35 days. As depicted in
Fig.~\ref{fig:flux_E_and_r}, this experiment ought to be capable of detecting the line-spectrum neutrinos from both the pep reaction and the high energy \(^7\text{Be}\) decay, as well as 
the continuous-spectrum neutrinos from \(^8\text{B}\) decay and the hep reaction. However, since the \(^{37}\text{Cl}\) ground state prefers to transit to the \(^{37}\text{Ar}\) excited states at an energy level of about 5 MeV above it, the Homestake experiment 
only detected the last two solar neutrinos, among which the \(^8\text{B}\) neutrinos were predominant. 

Another neutrino capture process, $\nu_e +{^{71}\text{Ga}}\rightarrow ~{^{71}\text{Ge}} + e^-$, has also been utilized by radiochemical experiments such as GALLEX/GNO and SAGE.
The energy threshold of this process is 233 keV,  lower than the end-point of 420 keV for the pp reaction, enabling them to access solar neutrinos from all sources. The daughter nucleus \(^{71}\text{Ge}\) has a half-life of 11.4 days, also reasonably long for chemical extraction. 

In addition, other possible nuclear isotopes have also received attention, including $^{98}\text{Mo}$, $^{203}\text{Tl}$, $^{7}\text{Li}$, 
$^{81}\text{Br}$, and $^{127}\text{I}$.
The chemical or geochemical experiments based 
on \(^{98}\text{Mo}\) and \( ^{203}\text{Tl}\) are sensitive to solar neutrino fluxes averaged over 
millions of years. If the SSM is correct, the average 
fluxes should be the same as the contemporary ones, providing a test to the SSM.
For those based on $^{7}\text{Li}$,
$^{81}\text{Br}$, and $^{127}\text{I}$, they are also similar to the detector based on \(^{37}\)Cl but with different energy thresholds, which are 862, 470, and 789 keV, respectively.  

It is worth mentioning that experiments based on the neutrino capture reactions can provide direct measurements of the neutrino energy spectrum if the energy of the final-state electron $E_e$ can be measured. The neutrino energy can be determined via simple kinematics:
\(E_\nu = M_\text{p}-M_\text{d}+E_e\), 
where \(M_\text{p}\) and \(M_\text{d}\) are the masses of parent and daughter nuclei, respectively.
If the daughter nucleus can provide a detectable delay signal correlating with the prompt electron signal in space and time, then they can be detected without traditional radiochemical means and the background can also be significantly suppressed. 
A few isotopes including $^{176}{\rm Yb}$, $^{160}{\rm Gd}$, and $^{82}\text{Se}$ have been proposed for solar neutrino detection with real-time signatures for 
discriminating solar neutrino signals from the potential radioactive background~\cite{Raghavan:1997ad}.

\subsection{Cherenkov experiments}

Cherenkov detectors for solar neutrino detection use water ($\text{H}_2\text{O}$) or heavy water ($\text{D}_2\text{O}$) as the target material.

\begin{figure}
	\centering 
	\includegraphics[width=0.5\columnwidth]{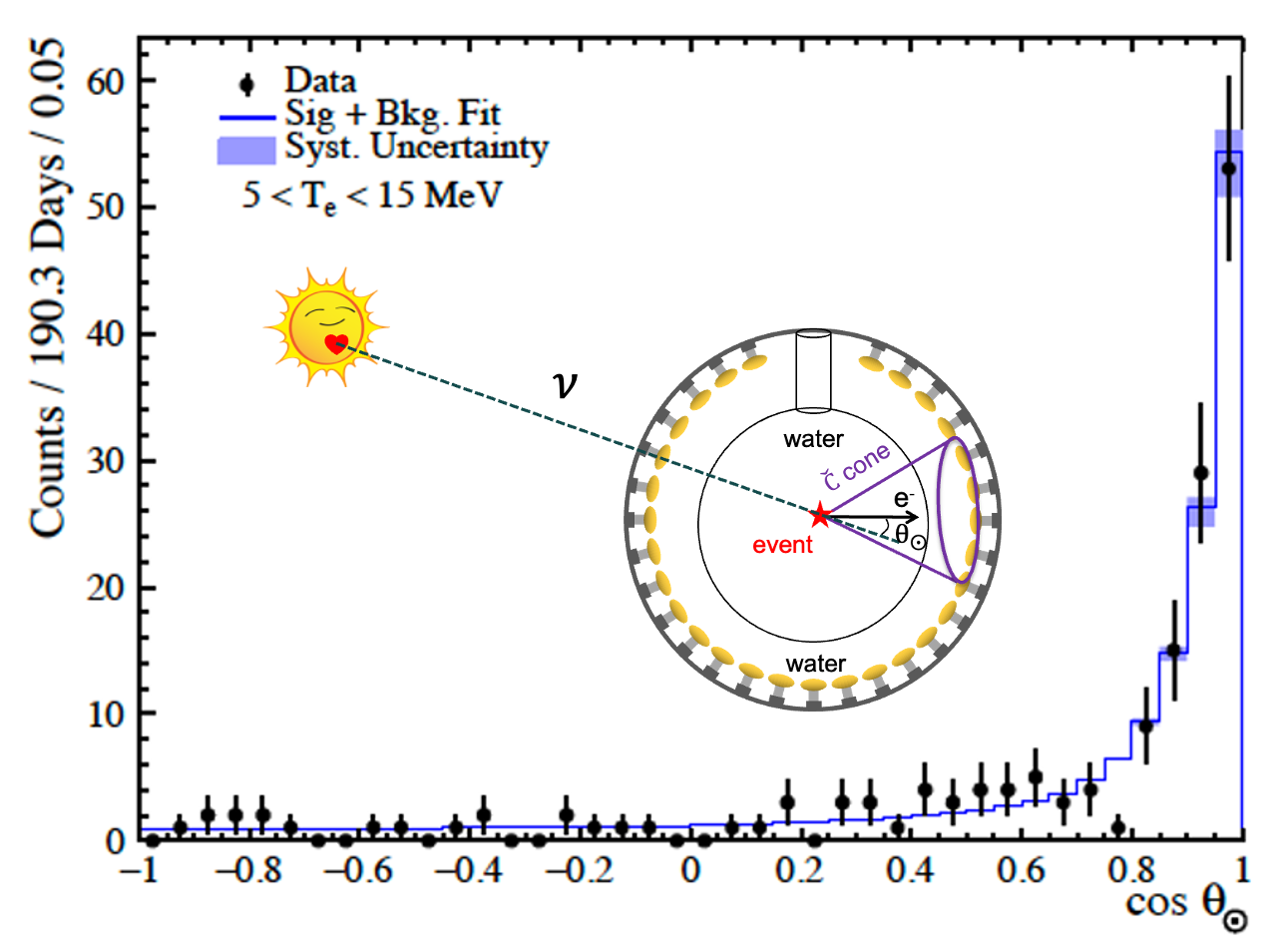} 
	\caption{Angular distribution of recoil electron in the SNO+ experiment, with the kinetic energy in 5-15 MeV.  Below 5 MeV, natural radioactivity causes significant backgrounds. }
		\label{fig:exp:solar_angle}
\end{figure}

In water-based Cherenkov detectors (such as Kamiokande and Super-Kamiokande), solar neutrinos are detected via elastic electron scattering,
\(\nu_{\alpha}+e^-\rightarrow~\nu_{\alpha}+e^-\), where the flavor index $\alpha$ can be any of $e$, $\mu$, and $\tau$.
The final-state $e^-$ emits Cherenkov light in the detector if its velocity is higher than the speed of light in the medium.  This makes both the angle and energy of the electron detectable.
It is noteworthy that the cross section of elastic $\nu_e+e^-$ scattering is significantly larger than that of $\nu_\mu+e^-$ or $\nu_\tau+e^-$. 

Due to the lightness of the electron, the angle between the final-state electron and incoming neutrino is generally small (for instance, a 6 MeV electron generated by a 10 MeV neutrino has the angle $\approx 14^{\circ}$), implying that the motion of the recoil electron is rather forward. 
This characteristic is exploited to determine the direction of incoming solar neutrinos (see Fig.~\ref{fig:exp:solar_angle}), effectively differentiating them from background sources such as environmental radiation and cosmogenic background. 
One of the advantages of water-based Cherenkov detectors is that they can be easily scaled up with real-time measurements and comparatively low cost on the target material (pure water). Therefore, water-based Cherenkov detectors typically have very high fiducial masses, ranging from kiloton to sub-megaton scales.

Heavy-water-based Cherenkov detectors can detect neutrinos via  $\nu_{e} + \text{D}_2\text{O} \rightarrow 2p + e^-$ and $ \nu_{\alpha} + \text{D}_2\text{O} \rightarrow \nu_{\alpha} + p + n$, in addition to elastic neutrino-electron scattering. 
The neutral-current processes $ \nu_{\alpha} + \text{D}_2\text{O} \rightarrow \nu_{\alpha} + p + n$ render heavy-water-based experiments more sensitive to all neutrino flavors. Historically, it played a crucial role in resolving the solar neutrino problem, as solar neutrinos that have changed flavors can be equally detected by this process. 

Since water (and heavy water) contains a large number of oxygen nuclei, the interactions between atmospheric neutrinos and oxygen nuclei, via \(\nu(\bar{\nu})+{{}^{16}\text{O}}\rightarrow~\nu(\bar{\nu})+n+{{}^{15}\text{O}^*}\) or 
\(\nu(\bar{\nu})+{{}^{16}\text{O}}\rightarrow~\nu(\bar{\nu})+p+{{}^{15}\text{N}^*}\), engender an unavoidable background for studying solar neutrinos. The subsequent de-excitation of the produced causes a gamma-ray background, which is indistinguishable from electron signals in Cherenkov detectors. Since this background is typically accompanied by the production of neutrons, neutron-tagging is important to the reduction of this background. 
The Super-Kamiokande experiment has developed a technique to dissolve gadolinium chemical compounds in water, notably enhancing neutron detection efficiency.  This advancement will potentially improve the rare signal search for hep neutrinos and solar antineutrinos.

\subsection{Liquid scintillator experiments}

Liquid scintillator (LS) experiments also employ elastic neutrino-electron scattering as the primary detection process for solar neutrinos, but
have much improved energy resolution compared to water- or heavy-water-based Cherenkov experiments. 
In LS,  neutrino-electron scattering is detected via the ionization energy deposited by the recoil electron. Since the electron deposits most of its kinetic energy in the form of ionization instead of Cherenkov radiation, the light yield in LS is much higher than that in water. 

LS is usually composed of organic solvent doped with fluorophores. The ionization caused by a charged particle excites the aromatic solvent molecules, which then transfer their excited energy to fluorophore molecules. Due to a phenomenon called the Stokes shift, the fluorophore emits photons with significantly larger wavelengths than the solvent's photon spectrum when it undergoes de-excitation, falling into the sensitive range for photon sensors or phototubes. 
As a result, LS can precisely measure the electron kinetic energy deposited in ionization by converting it to optical photons.
This enhanced capability is particularly important for detecting solar neutrinos, which have energies on the MeV scale. 

As a unpolarized medium, LS is easier to purify than water, making it more suitable for detecting pp, pep, \(^7\)Be, and CNO neutrinos. The Borexino experiment is a successful example of utilizing LS to detect solar neutrinos including these low-energy components.  LS is relatively cost-effective and can be scaled up as real-time detectors.

Recently, techniques for directional reconstruction utilizing the faint Cherenkov light emitted by electrons in liquid scintillators have been developed. One method involves extending the florescent decay time to give prominence to the Cherenkov light. This method has been successfully demonstrated in the SNO+ experiment. The other method utilizes a correlated and integrated directionality approach. By analyzing the detected phototube hit pattern in relation to the known position of the Sun and integrating it over large statistics of events, it is possible to generate a distribution of the angle between the hit phototubes and the interaction position of a solar neutrino with the matter. Electrons scattered by solar neutrinos create a distinct signature in the angular distribution compared to the isotropic radioactive background unrelated to the Sun. This signature allows for the separation and measurement of the solar neutrino signal. This technique has been effectively employed in the Borexino experiment, significantly enhancing its sensitivity in measuring the pep, $^7$Be, and CNO fluxes, especially in discovering CNO neutrinos.

\subsection{The solar neutrino problem}
The first measurement of solar neutrinos came out in 1968~\cite{Davis:1968cp} from the Homestake experiment led by Davis. 
The result was that the solar neutrino ﬂux was less than 3 SNU (Solar Neutrino Unit. One SNU corresponds to one capture per second per $10^{36}$ target nuclei). 
This result was significantly lower than the theoretical values calculated by John Bahcall~\cite{Bahcall:1968hc} (the theoretical values are 21, 11, 7.7, 4.4, and 11 SNU for five models considered by Bahcall in this paper).  Further data taking at the Homestake experiment led to a more precise measurement: $2.56\pm0.16\pm0.16$~SNU~\cite{Cleveland:1998nv},
which was only one-third of Bahcall's refined prediction~\cite{Bahcall:1987jc}.
Other neutrino experiments based on \(^{71}\)Ga and water targets also found deficits, though they varied from half to two-thirds. 
The discrepancy is known as the solar neutrino problem ~\cite{Bahcall:1976zz}.

For a decade or more after the Homestake experiment published the first result, the most common explanation for the problem was that something was incorrect with the solar model. Theories endeavored to amend solar models by incorporating additional effects into the Sun, such as magnetic fields, rapid rotation, and atypical metal abundances. However, as helioseismology was established, it became evident that there was little wrong with the solar model. Thus, the solution to the solar neutrino problem must lie in some new physics of elementary particles.

Although nowadays we know the fundamental cause of the solar neutrino problem is neutrino oscillation, it is nevertheless worth mentioning a few historical attempts to address this problem. In an article titled ``What cooks with solar neutrinos?''~\cite{Fowler:1972zz}, Fowler proposed 
two explanations involving experimental nuclear physics, and changing theoretical solar structure and evolution.
Cisneros followed a discussion with Werntz and investigated an extreme assumption that a neutrino could alter its helicity state when passing through the strong magnetic field inside the Sun and change to an antineutrino~\cite{Cisneros:1970nq}.
Bahall, Cabibbo, and Yahil raised the question of whether neutrinos were stable particles and discussed the consequence of neutrinos having finite masses~\cite{Bahcall:1972my}. 
Freedman et al. suggested using a mass-spectrometric assay of the induced tiny concentration of \(1.6\times10^7\) \(^{205}\)Pb in old thallium minerals to examine the dominant low-energy component in solar neutrino flux~\cite{Freedman:1976exx}. 
Bahcall et al. proposed to use \(^{71}\)Ga to trap lower-energy solar neutrinos~\cite{Bahcall:1978qq}. 
Haxton and Cowan proposed studying long-lived isotopes produced by solar neutrinos in the Earth's crust to probe secular variations in the rate of energy production in the sun's core~\cite{Haxton:1980wg}. 
Faulkner and Gilliland hypothesized an idea, assuming that a small mass fraction of weakly interacting massive particles (WIMPs) lurked in the core of the sun and served as very efficient energy conductors, which could change the core temperature and affect solar neutrino production~\cite{Faulkner:1985rm}.  

Eventually, neutrino oscillation turned out to be right answer to the solar neutrino problem. The theory of neutrino oscillation will be introduced in the next section.

\section{Solar neutrino oscillation \label{sec:osc}}

The idea of neutrino oscillation was first proposed by Pontecorvo in the 1950s with the original consideration on $\nu \leftrightarrow \overline{\nu}$ oscillation and later developed to incorporate flavor mixing and the matter effect. The modern theory of neutrino oscillation is formulated for the purpose of computing flavor transitions among three neutrino flavors ($\nu_e$, $\nu_{\mu}$, $\nu_{\tau}$), applicable to neutrino oscillation in various circumstances including the Sun. In this section, we briefly review the modern theory of neutrino oscillation and a few well-known phenomena in solar neutrino oscillation.

\subsection{The general formalism for neutrino oscillation}
If neutrinos have masses, their mass eigenstates ($\nu_i$ with $i=1,2,3, \cdots$) are not necessarily in alignment with their flavor eigenstates  ($\nu_{\alpha}$ with $\alpha=e,{\mu},\tau, \cdots$)  participating in weak interactions. The two sets of eigenstates may differ by a unitary transformation, $\nu_\alpha=\sum_i U_{\alpha i} \nu_i$, where $U$ is the so-called PMNS matrix.

Neurino oscillation refers to the phenomenon that neutrinos in a specific flavor eigenstate, which is a quantum superposition of mass eigenstates, may change to another flavor eigenstate during propagation, due to the different dispersion relations of mass eigenstates.
Quantitatively, the evolution of neutrino flavors in the three-neutrino framework of neutrino oscillation is governed by the following equation:
\begin{equation}
	i\frac{d}{dL}\left(\begin{array}{c}
	\nu_{e}\\
	\nu_{\mu}\\
	\nu_{\tau}
	\end{array}\right)=\left[\frac{1}{2E_{\nu}}U\left(\begin{array}{ccc}
	m_{1}^{2}\\
	 & m_{2}^{2}\\
	 &  & m_{3}^{2}
	\end{array}\right)U^{\dagger}+\left(\begin{array}{ccc}
	V_{e}\\
	 & 0\\
	 &  & 0
	\end{array}\right)\right]\left(\begin{array}{c}
	\nu_{e}\\
	\nu_{\mu}\\
	\nu_{\tau}
	\end{array}\right)\thinspace,\label{eq:sch}
\end{equation}
where $L$ denotes the propagation distance, $E_{\nu}$ is the neutrino
energy, $m_{1,2,3}$ are the masses of $\nu_{1,2,3}$, and $V_{e}=\sqrt{2}G_{F}n_{e}$
is an effective potential with $G_{F}$ the Fermi constant and $n_{e}$
the electron number density of the medium.

The effective potential $V_{e}$ accounts for the matter effect, also
known as the Mikheyev-Smirnov-Wolfenstein (MSW) effect~\cite{Wolfenstein:1977ue,Mikheev:1986gs,Mikheev:1986wj}, on neutrino
oscillation. It is caused by coherent forward scattering of neutrinos
with medium particles. In principle, both electrons and nuclei contribute
to the effective potential but the contribution of the latter is flavor
independent because it is caused by flavor-blind neutral-current interactions. Such a flavor-independent contribution does not affect oscillation and can be neglected. 

\subsection{The survival probability of solar electron neutrinos\label{sub:MSW}}

Eq.~\eqref{eq:sch} can be straightforwardly applied to solar neutrino
oscillation. Both the solar and terrestrial matter effects can be
readily taken into account by including $L$-dependent contributions to $V_{e}$. In practice, we are mainly concerned with the survival probability,
$P_{ee}\equiv\left|\langle\nu_{e}(L)|\nu_{e}(0)\rangle\right|^{2}$,
which is the probability of an electron neutrino produced at the source
(denoted by $|\nu_{e}(0)\rangle$) after traveling through the distance
$L$ still retaining the original flavor (i.e.~in the state $|\nu_{e}(L)\rangle$).

\begin{figure}
	\centering 
	\includegraphics[width=0.6\textwidth]{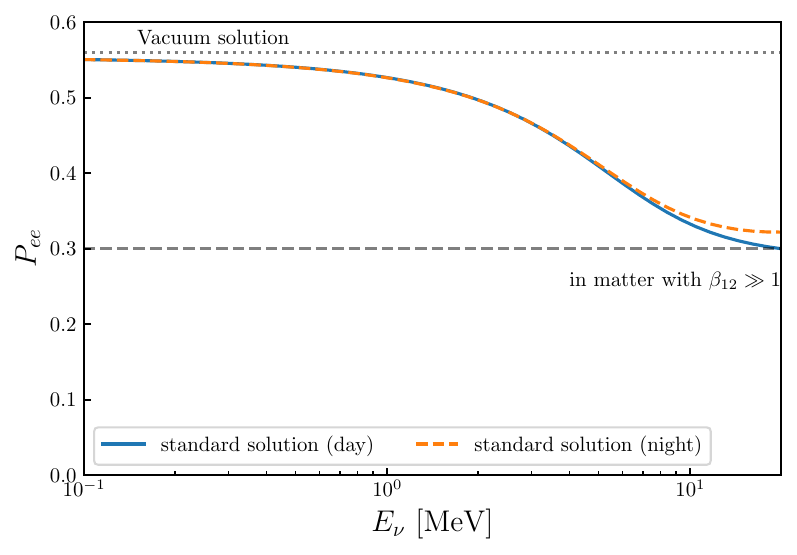} 
	\caption{The probability of a solar neutrino retaining its original flavor ($\nu_e$) after arriving at the Earth, known as the survival probability. The survival probability approaches the vacuum solution (about 0.57) in the low-energy limit, and decreases to lower values at high energies due to the matter effect. The  matter effect includes contributions not only from the solar medium but also from the terrestrial matter, as is indicated by the difference between solid and dashed lines.
	\label{fig:exp:P_survival}} 
\end{figure}

Under certain approximations, one can compute $P_{ee}$ analytically
without numerically solving Eq.~(\ref{eq:sch}). Assuming that the
evolution is adiabatic (which means $V_{e}$ varies sufficiently slowly
in the Sun compared to the oscillation wavelength) and $\nu_{3}$
is not involved in the oscillation ($\nu_{e}$ mainly consists of
$\nu_{1}$ and $\nu_{2}$), $P_{ee}$ is given by 
\begin{equation}
P_{ee}\approx\frac{1}{2}+\frac{1}{2}\cos2\theta_{12}^{m}\cos2\theta_{12},\label{eq:p-ee}
\end{equation}
with 
\begin{align}
\cos2\theta_{12}^{m} & \approx\frac{\cos2\theta_{12}-\beta_{12}}{\sqrt{(\cos2\theta_{12}-\beta_{12})^{2}+\sin^{2}2\theta_{12}}}\thinspace,\ \ \beta_{12}\equiv\frac{2V_{e}^{0}E_{\nu}}{\Delta m_{21}^{2}}\thinspace,\label{eq:theta-m}
\end{align}
where $\theta_{12}$ is an angle quantifying the composition of $\nu_{e}$
($\nu_{e}\approx\cos\theta_{12}\nu_{1}+\sin\theta_{12}\nu_{2}$),
$\Delta m_{21}^{2}\equiv m_{2}^{2}-m_{1}^{2}$, and $V_{e}^{0}$ denotes
the value of $V_{e}$ at the solar center. Taking specific values
of $V_{e}^{0}$ and $\Delta m_{21}^{2}$, $\beta_{12}\approx E_{\nu}/(3\ \text{MeV})$.
Note that $P_{ee}$ is energy dependent. Figure \ref{fig:exp:P_survival}
shows the variation of $P_{ee}$ as a function of $E_{\nu}$. 

There are two interesting limits of the survival probability. At low
energies ($E_{\nu}\ll3$ MeV), $\beta_{12}$ in Eq.~\eqref{eq:theta-m}
can be neglected, leading to $\theta_{12}^{m}\to\theta_{12}$ and
$P_{ee}\approx(1+\cos^{2}2\theta_{12})/2\approx0.57$. This is known
as the vacuum limit since $P_{ee}$ in this limit is almost unaffected
by the MSW effect. At high energies ($E_{\nu}\gg3$ MeV), Eq.~\eqref{eq:theta-m}
gives $\cos2\theta_{12}^{m}\approx-1$ and $\theta_{12}^{m}\approx90^{\circ}$,
implying that the neutrino coming out of the Sun would be almost purely
$\nu_{2}$. In this high-energy limit, the survival probability is
given by $P_{ee}\approx(1-\cos2\theta_{12})/2\approx\sin^{2}\theta_{12}\approx0.3$,
corresponding to the gray dashed curve in Fig.~\ref{fig:exp:P_survival}. 
The transition from the high-energy to the low-energy limits, occurring at around a few MeV, is referred to as the \emph{upturn} in the literature.

When solar neutrinos arrive at a detector on the Earth at night, the
neutrinos have also traversed a significant length of terrestrial matter.
This causes the survival probability in the high-energy limit to be
slightly higher than that without the earth matter effect. Hence it
is fair to say that ``the Sun at night is brighter than that in the
day'' if the brightness refers to the luminosity of $\nu_{e}$. More
specifically, the day-night difference of $P_{ee}$ can be estimated
by
\begin{equation}
\delta P_{ee}\equiv P_{ee}^{(\text{day})}-P_{ee}^{(\text{night})}\approx\frac{1}{2}\frac{\cos2\theta_{12}^{m}\sin^{2}2\theta_{12}\beta_{\oplus}}{\beta_{\oplus}^{2}-2\beta_{\oplus}\cos2\theta_{12}+1}\thinspace,\label{eq:day-night}
\end{equation}
where $\beta_{\oplus}$ is similar to $\beta_{12}$ in Eq.~\eqref{eq:theta-m}
except that $V_{e}^{0}$ is replaced by the average value of $V_{e}$
in the Earth. In Fig.~\ref{fig:exp:P_survival}, the small difference
between the orange dashed and blue solid lines represents this correction due to the Earth matter effect.

\section{Experimental achievements and progress \label{sec:exp-progress}}
Since the first solar neutrino experiment, many experiments have made important progress, advancing our understanding of the Sun and  
neutrinos. 
In this section, we briefly review the achievements of measuring solar neutrino fluxes.

\subsection{Measurements of solar neutrino fluxes}

According to the standard solar model, solar neutrinos contains
$^8$B, $^7$Be, pep, pp, CNO, and hep neutrino components. 
Almost all these components have been observed except for the hep neutrinos. Below we briefly review the measurement of each component. 

\subsubsection{$^8{\rm B}$ neutrinos}
Solar $^8$B neutrinos are produced from the decay of $^8$B, which is a product of the pp-chain reactions, as is shown Fig.~\ref{fig:pp-CNO}. 
The maximum energy of $^8$B neutrinos is around 15 MeV. 
From Fig.~\ref{fig:flux_E_and_r} one can see that $^8$B neutrinos have the highest flux in the energy range from $3$ to $15$ MeV, which makes $^8$B neutrinos the most successfully measured component.

Currently, the Super-Kamiokande experiment provides 
the most precise flux measurement of $^8$B neutrinos using neutrino-electron scattering. 
Assuming all $^8$B neutrinos are in the $\nu_e$ state, the measure flux is
\((2.336\pm0.011\pm0.043)\times 10^6~\text{cm}^{-2}\text{s}^{-1}\), which significantly deviates from the SSM
prediction. According to the survival probability in Fig.~\ref{fig:exp:P_survival}, this deficit is expected from neutrino oscillation.

Note that the above measurement is sensitive to neutrino oscillation parameters. 
It is possible to measure the $^8$B neutrino flux without being affected by neutrino oscillation. This is achieved by the SNO experiment using heavy water as the target material.
As mentioned in Sec.~\ref{sec:exp}, heavy water allows the pure NC process $ \nu_{\alpha} + \text{D}_2\text{O} \rightarrow \nu_{\alpha} + p + n$ to be exploited for neutrino detection. This process is flavor-independent so the result is invulnerable to the uncertainties of oscillation parameters.
In addition,  the CC process $\nu_{e} + \text{D}_2\text{O} \rightarrow 2p + e^-$ also offer a $\nu_e$-only measurement of the flux. 
From 1999 to 2006, the SNO experiment underwent three phases, using distinct neutron detection methods to improve the flux measurement via the NC process and  the $\nu_e$ flux measurement via the CC process. 
A combination of all three phases of data gives the final result  
\((5.25\pm0.16 ^{+0.11}_{-0.13})\times 10^6~ \text{cm}^{-2}\text{s}^{-1}
\), with an uncertainty of around 3.8\%. 
JUNO, an upcoming 20-kiloton liquid scintillator experiment, will explore the feasibility of using isotope $^{13}$C with a natural abundance of 1.1\% to detect solar neutrinos, and may provide another model-independent flux measurement. 

\subsubsection{$^7{\rm Be}$ neutrinos}

$^7$Be neutrinos are produced via the electron capture process
\(
^7\text{Be}+e^-\rightarrow~^7\text{Li}+\nu_e
\). The energy spectrum contains two monochromatic lines, one at 0.861 MeV (90\%) and the other at 0.383 MeV (10\%).

The experimental study of sub-MeV solar neutrinos is challenging. It requires both a clean environment with extremely low radioactivity and an excellent detector with high energy resolution. 
The Borexino experiment was designed to detect the 0.862 MeV line in a real-time detector with a low natural radioactive background and high energy resolution. 
So far, only the Borexino experiment has successfully measured the $^7$Be flux. The result is
$\Phi(^7\text{Be})=(4.99\pm0.11^{+0.06}_{-0.08})\times 10^9 \text{cm}^{-2}\text{s}^{-1}$, in agreement with the expected value given in Table~\ref{tab:flux}. 

The Borexino experiment employed neutrino-electron scattering as the detection process. 
This actually only allows the electron energy instead of the neutrino energy to be  measured.  
In the sub-MeV energy range, the electron recoil spectrum of $^7$Be neutrinos  overlaps with those of other solar neutrino components. So one has to perform a spectral fit to extract the $^7$Be component, assuming the standard oscillation with known oscillation parameters.  Note that the energy of $^7$Be neutrinos lies in the transition phase between vacuum oscillation and matter effect---see Fig.~\ref{fig:exp:P_survival}. Since this part is known to be sensitive to new physics effects, it would be desirable that future solar neutrino experiments can measure the neutrino energy directly (i.e. serving as a neutrino energy spectroscopy).

\subsubsection{pp neutrinos}
The fundamental thermonuclear reaction producing solar energy is the proton-proton reaction
\(
    p+p\rightarrow~^2\text{H}+e^++\nu_e, 
\)
with a maximum neutrino energy of 0.42 MeV.

The pp neutrino flux constitutes 91\% of the total solar neutrino flux, making it the dominant component. 
However, it also has the lowest energy, which makes its detection particularly challenging, similar to the situation of $^7$Be neutrinos.
The Borexino experiment has, thanks to the excellent performance of its liquid scintillator and the low background,  successfully measured the flux of pp neutrinos: 
\(\Phi(\text{pp})=(6.1\pm0.5^{+0.3}_{-0.5})\times 10^{10} ~\text{cm}^{-2}\text{s}^{-1}\), which is consistent with the SSM predictions in Table~\ref{tab:flux}. 
The measurement of pp neutrinos at Borexino is achieved by collecting neutrino-electron scattering data with a very low detection threshold down to 0.165 MeV. 
Due to the overlapping with the other solar neutrinos, the measurement of the pp-neutrinos is obtained via a spectral fit.
So far, the precision of pp neutrino  measurement still cannot discriminate the solar models listed in Table~\ref{tab:flux}.

\subsubsection{pep neutrinos}

In proton-proton fusion, there is a small probability (0.24\%) that an electron is captured by a proton before the fusion, leading to the pep process:
\(
    p+e^-+p\rightarrow~^2\text{H}+\nu_e
\).
The two-particle final state makes the neutrino energy monoenergetic, with the neutrino energy at 1.44 MeV. This energy lies in the transition phase from vacuum oscillation to matter effect and, therefore, is important for testing various oscillation models. The first evidence of the pep solar neutrinos was directly detected by the Borexino experiment. Even though 
it has a monoenergetic feature, 
extracting the pep neutrino component from the electron energy spectrum is complicated since its energy distribution overlaps with the one for the CNO neutrinos. In addition to  
assuming the Mikheyev-Smirnov-Wolfenstein large mixing angle solution to solar neutrino oscillations, 
one has to fix the shape of the CNO 
neutrinos, which depends on whether High-Z SSM or Low-Z SSM metallicity is used for the CNO neutrinos. Consequently, the flux is determined with the two models:
\(
\Phi(\text{pep})=(1.27\pm0.19^{+0.08}_{-0.12})\times 10^8 \text{cm}^{-2}\text{s}^{-1}
\) for High-Z SSM metallicity,
\(
\Phi(\text{pep})=(1.39\pm0.19^{+0.08}_{-0.13})\times 10^8 \text{cm}^{-2}\text{s}^{-1}
\) for Low-Z SSM metallicity, respectively. Both results are consistent with the expected value in Table~\ref{tab:flux}. 

\subsubsection{CNO neutrinos}

The sub-dominant CNO cycle involves fusion facilitated by the presence of carbon, nitrogen, and oxygen. The rates of these processes are dependent on temperature. The CNO cycle is comprised of two sub-cycles, CN and NO. At the relatively low temperature of the solar core, sub-cycle CN is the primary process, accounting for about 99\% and producing neutrinos from the beta decays of $^{15}$O and $^{13}$N. The fusion facilitated by carbon, nitrogen, and oxygen provides valuable information regarding the metallicity of the Sun’s core, specifically, its abundance of elements heavier than helium. The solar metallicity in standard solar models leads to significantly different predictions for the CNO neutrino flux. A precise measurement serves as a crucial test to these models. The measurement was challenging due to the energy lying in the range of a few MeV, where radioactive and cosmogenic backgrounds are dominant. Thanks to the successful development of a technique for correlated and integrated directionality for sub-MeV solar neutrinos, Borexino ultimately discovered the CNO neutrinos and measured the flux to be 
\(
\Phi(\text{CNO})=(6.7^{+1.2}_{-0.8})\times 10^{8} \text{cm}^{-2}\text{s}^{-1}
\).
This measurement aligns with the expected value, as provided in Table~\ref{tab:flux}. It is consistent with high metallicity standard solar models. When combined with the flux measurements on $^8$B and $^7$Be, the low metallicity SSM is disfavored, offering direct experimental access to the study of the primary mechanism for the conversion of hydrogen into helium in the Universe.

\subsubsection{hep neutrinos}

The last and most difficult process through which neutrinos are produced in the Sun is the fusion of protons and helium nuclei:
\(
    ^3\text{He}+p\rightarrow~^4\text{He}+e^++\nu_e
\).
This process is known as the hep-branch of proton-proton fusion. The neutrinos created by this process are called hep neutrinos. They have the highest energy (18.77 MeV) among all the solar neutrinos. Due to their small quantity and the end-point energy slightly above the $^8$B neutrinos, the search for hep neutrinos is somewhat tricky. It is now the only one that solar neutrino experiments have not yet detected. Attempts have been made in both the Super-Kamiokande and the SNO experiments, but no evidence has been found. The current strict limit is from the SNO experiment:
\(
\Phi(\text{hep})< 2.3\times 10^{4} \text{cm}^{-2}\text{s}^{-1}~@90\%~\text{C.L.}
\).
Based on a comparison to the expected value provided in Table~\ref{tab:flux}, it is evident that significant improvements are necessary to detect them.
Discovering the hep neutrinos would greatly impact astroparticle physics, particularly in our comprehension of stellar evolution and the physics of massive neutrinos, as this process in the Sun's fusion core drives a much higher neutrino production than any other reaction.

\subsection{Solar neutrino oscillation parameters}

Two types of solutions exist for the observed neutrino missing problem. Either the solar neutrino oscillates in vacuum during its journey to the Earth, or it has already been converted within the Sun due to matter effects. 

In the vacuum oscillation solution, as previously mentioned, considering the observable \(^8\)B neutrino ratio relative to the expected from the SSM, which is about 0.45, one can use the vacuum oscillation formula to estimate the neutrino mass-square difference regions of around \(10^{-10}~\rm{eV}^2\). 
However, assuming CPT invariant, this value is inconsistent with the observation of the KamLAND experiment using $\bar{\nu}_e$'s from commercial nuclear reactors in a similar energy range but with a shorter distance, about 180 km, so short that the oscillation can be treated in vacuum. 
From a study of the energy spectrum, the KamLAND experiment yields
\(\rm{sin}^2_{\theta_{12}}=
0.325^{+0.062}_{-0.054}\), 
and
\(\Delta m^2_{21}=(7.54^{+0.19}_{-0.18})
\times 10^{-5}\rm{eV}^2\). 
Consequently, the solution using neutrino oscillation in vacuum can be excluded to explain the deficit observed in measuring the 
high-energy \(^8\)B neutrino flux. 

The other appealing solution is a conversion in matter via the MSW effect. In this scenario, 
electron neutrinos scattering off electrons in the high-density Sun interior can cause the almost complete conversion of $\nu_e$'s to $\nu_\mu$'s and $\nu_\tau$'s. 
The Super-Kamiokande and the SNO experiments use solar neutrino data to provide a combined fit to the three-flavor neutrino oscillation parameters, which are
\(\rm{sin}^2_{\theta_{12}}=
0.305\pm0.014\), and
\(\Delta m^2_{21}=(6.10^{+1.04}_{-0.75})
\times 10^{-5}\rm{eV}^2\). 
These results can be compared with those from the KamLAND experiment. 
Both results are consistent, though solar experiments have a larger error in the mass-square difference due to the lack of data covering a few MeV regions, where solar neutrinos undergo a transition from vacuum oscillation to matter effect. It is expected that the JUNO experiment will significantly improve the precision using reactor antineutrinos. 

\subsection{Terrestrial matter effects}

The cleanest and most direct test for terrestrial matter effects on neutrino oscillations lies in the comparison of the daytime and the nighttime solar neutrino interaction rates. In this comparison, the solar zenith angle, defined as the angle between the vector from the solar position to the solar neutrino event position and the vertical detector (z) axis, governs the size and length of the terrestrial matter density through which the neutrinos pass and, thereby, the oscillation probability and the observed interaction rate. An increase in the nighttime interaction rates implies a regeneration of electron-flavor neutrinos.
A day/night asymmetry parameter is defined as
\[
A_{\rm{D/N}}=2\frac{\Phi{^{\rm{day}}}
-\Phi{^{\rm{night}}}}
{\Phi{^{\rm{day}}}
+\Phi{^{\rm{night}}}}\,.
\]
The expected asymmetry also depends on the oscillation parameters and the energy range within which the flux is measured. With current oscillation parameters, the day/night asymmetry is anticipated to be at a few percent level in the MeV and above region. In contrast, no asymmetry is expected in the keV region. The SNO experiment employs the $^8$B neutrino events from the ES, CC, and NC channels to measure the day/night asymmetry, consistent with zero. Although the Borexino experiment utilizes the \(^7\)Be neutrino events to measure the asymmetry, its result is also consistent with zero. By contrast, the Super-Kaomiokande experiment uses the 
large dataset of 
\(^8\)B neutrino events to determine the day/night asymmetry parameter, giving
\(A^{\rm{SK}}_{\rm{D/N}}=-0.0286\pm0.0085\pm0.0032
\),
which deviates from zero by \(3.2\sigma\), providing evidence for the existence of earth matter effects on solar neutrino oscillation, namely
the terrestrial matter effects. It is expected 
that the successor of the Super-Kamiokande experiment, the Hyper-Kamiokande experiment, will be capable of increasing the sensitivity up to \(5\sigma\) and will eventually discover the terrestrial matter effects. 

\subsection{Solar activity and seasonal effect}

Owing to the high-energy characteristic of the 
\(^8\)B neutrino events, experiments can offer a real-time clean and highly statistical flux measurement for studying the potential correlation between the annual sunspot activity and the \(^8\)B neutrino flux. The measurement reveals a constant solar neutrino flux emitted by the Sun, at least for \(^8\)B neutrinos. It implies that the periodic activities of the Sun, such as the rotation within the Sun or the fluctuation of the sunspot numbers, have no impact on the thermonuclear reactions at the core of the Sun. The seasonal variation of \(^8\)B  flux is also utilized to examine the law of the inverse square of the distance between the Sun and the Earth, and no deviation is observed.

\section{Beyond the standard framework \label{sec:beyond}}  

When the solar neutrino problem appeared, many hypotheses were proposed to resolve the problem. After tremendous experimental efforts, the standard framework of neutrino oscillation has been established and accepted insofar as the only successful resolution. Nevertheless, new physics beyond the standard framework might lurk in solar neutrinos and could be revealed by precision measurements of solar neutrino fluxes. Below, we briefly mention a few of such possibilities.

\subsection{Sterile neutrinos}
Sterile neutrinos, as suggested by the name, are neutrino-like particles possessing two features: (i) they do not participate in SM gauge interactions, and  (ii) they have mass mixing with the SM neutrinos. By contrast, the SM neutrinos in this context are often referred to as active neutrinos. 
In the presence of light sterile neutrinos, solar neutrino oscillation may contain new oscillation modes, and the corresponding phenomenology has been widely discussed in the literature. In particular, it has been shown that sterile neutrinos could modify the vacuum-matter transition (known as the up-turn) of the survival probability in Fig.~\ref{fig:exp:P_survival}, causing a significant dip on the curve at a few MeV. Therefore, precision measurements of the solar neutrino spectrum in this energy range are crucial to sterile neutrino searches. 

\subsection{Non-standard interactions}

In addition to the weak interactions predicted by the SM, many models proposed for massive neutrinos also suggest the existence of new neutrino interactions. A large collection of them can be parametrized in the framework called Non-Standard Interactions (NSIs). NSIs are four-fermion effective interactions similar to the Fermi interactions but with more general flavor structures. 

NSIs have two effects on solar neutrinos: (i) they can affect neutrino propagation in the solar medium, and (ii) they may modify the cross section of neutrino scattering at detection. The first effect was already considered by Wolfenstein in his seminal paper on the MSW effect, where it was suggested that even if neutrinos were massless, neutrino oscillation could still be induced by flavor off-diagonal NSIs. 
The second effect directly alters event rates in detectors. Therefore, when the relevant oscillation parameters are well determined, precision measurements of solar neutrino event rates can be used to set stringent constraints on NSI parameters.

\subsection{Other new physics scenarios}
In addition, many other new physics scenarios, such as neutrino magnetic moments,  spin-flavor precession, neutrino decay, light mediators, and WIMPs, could potentially influence solar neutrino observations. Interested readers are referred to Ref.~\cite{Xu:2022wcq} for a more comprehensive review of this subject.

\section{Summary and outlook\label{sec:summary}}

The study of solar neutrinos began with the unexpected deficit in observations known as the solar neutrino problem. This issue prompted numerous hypotheses and tests, ultimately culminating in the groundbreaking discovery of neutrino oscillation. After more than half a century of research, the major solar neutrino fluxes have been accurately measured and are in good agreement with predictions from the standard solar model combined with the interpretation of neutrino oscillation.

However, solar neutrino physics is far from approaching the end. The standard solar models still face unresolved challenges, such as the metallicity problem, and there may be new physics lurking within the properties of neutrinos. With the advent of new experiments, solar neutrino research is entering an exciting new era.

Precision measurements of solar neutrino energy spectrum, fluxes, and neutrino-mixing parameters will provide deeper insights not only into solar physics itself but also into particle physics.  This field continues to serve as a unique probe for new physics. 
In the foreseeable future, next-generation experiments such as Hyper-Kamiokande, DUNE, and JUNO are anticipated to significantly enhance current measurements and present exciting opportunities for groundbreaking discoveries. As history has demonstrated, new observations of our nearest star may lead to profound advancements in our understanding of the fundamental laws of nature.

\begin{ack}[Acknowledgments]%
This work is supported in part by the National Natural Science Foundation of China under grant No.~12141501 and No.~12127808, and also by the CAS Project for Young Scientists in Basic Research (YSBR-099).
\end{ack}



\bibliographystyle{Numbered-Style} 
\bibliography{ref}

\end{document}